# Zener Tunneling Breakdown in Phase-Change Materials Revealed by Intense Terahertz Pulses

Yasuyuki Sanari[1], Takehiro Tachizaki[2,3], Yuta Saito[4], Kotaro Makino[4], Paul Fons[4], Alexander V. Kolobov[4], Junji Tominaga[4], Koichiro Tanaka[3,5], Yoshihiko Kanemitsu[1], Muneaki Hase[6,*], and Hideki Hirori[1,3,†]

[1]*Institute for Chemical Research, Kyoto University, Uji, Kyoto 611-0011, Japan*
[2]*Department of Optics and Imaging Science and Technology, Tokai University, Hiratsuka, Kanagawa 259-1292, Japan*
[3]*Institute for Integrated Cell-Material Sciences (iCeMS), Kyoto University, Kyoto, Kyoto 606-8501, Japan*
[4]*Nanoelectronics Research Institute, National Institute of Advanced Industrial Science and Technology, Tsukuba Central 5, 1-1-1 Higashi, Tsukuba 305-8565, Japan*
[5]*Department of Physics, Graduate School of Science, Kyoto University, Kyoto, Kyoto 606-8502, Japan*
[6]*Division of Applied Physics, Faculty of Pure and Applied Sciences, University of Tsukuba, 1-1-1 Tennodai, Tsukuba 305-8573, Japan*

**Abstract:** We have systematically investigated the spatial and temporal dynamics of crystallization that occurs in the phase-change material $Ge_2Sb_2Te_5$ upon irradiation with an intense terahertz (THz) pulse. THz-pump–optical-probe spectroscopy revealed that Zener tunneling induces a nonlinear increase in the conductivity of the crystalline phase. This fact causes the large enhancement of electric field associated with the THz pulses only at the edge of the crystallized area. The electric field concentrating in this area causes a temperature increase via Joule heating, which in turn leads to nanometer-scale crystal growth parallel to the field and the formation of filamentary conductive domains across the sample.

* mhase@bk.tsukuba.ac.jp, † hirori@scl.kyoto-u.ac.jp

Ge$_2$Sb$_2$Te$_5$ (GST) is one of the best performing chalcogenide alloys for phase-change applications, and its use in optical recording media owes to its robust optical contrast upon optical excitation inducing a phase change between amorphous and crystalline [1–3]. Because the phase change also enables resistive switching, which can be induced by nanosecond electrical pulses, GST is also suitable for electrical non-volatile memory applications [4,5]. It has been argued that GST sandwiched between two electrodes reaches its crystallization temperature through Joule heating. However, the electric field effects on resistive switching are still unclear, because many physical processes may simultaneously participate with both thermal and electric field effects. These processes range from simple atomic structural changes [2,6-11] to complex combinations of Joule heating-induced crystallization [4,12-14], nanometer-scale crystal growth or amorphization [15,16], nonlinear increases in electrical conductivity [17,18], and formation of filamentary paths across the active material between the electrodes [19,20].

Additionally, the phase changes induced by subnanosecond duration electrical pulses lead to a dome-shaped crystalline area [21,22], which is considered as a result of isotropic heat diffusion governing the spatial expansion of the crystal area. The experimental condition obscures the complex mechanisms of resistive switching [4,13,14], where the effects of both heating and electrical fields contribute to the phase change. To understand the relevant carrier dynamics, it is necessary to separate the contributions of these components by performing a sophisticated experiment. Recent advances in generating intense terahertz (THz) pulses have yielded new strategies for manipulating the electron and lattice degrees of freedom on picosecond time scales [23,24]. The use of ultrashort THz pulses helps to suppress heat diffusion and may enable the study of crystallization mechanisms in which the lattice temperature exceeds the crystallization temperature on picosecond time scales.

The present study used THz pulses of a few picoseconds to induce crystallization of amorphous GST. Pulses with fields up to a few MV/cm were generated through local field enhancement using a gold dipole antenna; the resulting fields were an order of magnitude higher than those in previous reports on electrical conductivity [25]. THz-pump–optical-probe spectroscopy combined with microscopy and micro-Raman measurements enabled a spatially resolved investigation of the effects of the electric field on the amorphous and crystalline GST areas within the antenna gap.

The experimental configuration is shown in Fig. 1(a) [26]. THz pulses were generated with the tilted pulse-front technique using a $LiNbO_3$ crystal and the setup described in Ref. [27,28]. The amorphous GST sample (40 nm thick) was deposited by sputtering a GST alloy target onto a (100) Si substrate at room temperature. A thin $ZnS-SiO_2$ layer (20 nm) was grown on top of the film in the same sputtering chamber to prevent oxidation. Figure 1(b) shows the temporal profile of the incident THz electric field. The field was enhanced by gold antennas fabricated on the sample surface. Figure 1(c) plots finite difference time domain (FDTD) calculations of the spatial distribution of the enhanced electric field and its direction (white arrows) inside the sample located in the 5-μm gap between the antennas and at a depth of 10 nm from the interface between the $ZnS-SiO_2$ and GST layers. The gap of the present sample is five times larger than that of a similar previous work [18], which allows us to trace the crystal growth much easier. The color bar shows the ratio between the maximum electric field of the incident pulse, $E_o$, and the electric field induced inside the sample, $E_i$. The enhancement factors were about five in the center of the gap and more than 15 near the antenna corners, giving $E_i$ values ranging from 1 to 3 MV/cm.

A THz pulse with a peak field of $E_o$ = 175 kV/cm (500 Hz repetition rate) was used to illuminate the sample, and the corresponding changes in reflectivity, $\Delta R/R$, were monitored by an optical microscope equipped with a charge coupled device (CCD) [Figs. 1(d)–1(g)]. The white areas in the images correspond to those with increased reflectivity in the visible light region. These figures show that changes begin to appear after about 10000 pulses, and an altered region with an elongated shape develops during the subsequent 5000 pulses. Because this growth process coincides with the direction of the electric field in Fig. 1(c), this phenomenon is due to the THz pulse irradiation [31].

We used Raman spectroscopy to identify the structure of the regions showing enhanced reflectivity. Figure 2(a) compares the spectrum of the region in Fig. 1(g) that became completely white (red) with the spectra for amorphous GST (blue) and crystalline GST (black); the latter crystalline sample was prepared by optical-pulse irradiation (about 1 mJ/cm$^2$, $\lambda$ = 800 nm) of the amorphous phase. The characteristic peaks of amorphous GST are at 130 and 148 cm$^{-1}$ ($A_1$ modes), while those of crystalline GST are at 105 and 160 cm$^{-1}$ ($E_g$ and $A_{1g}$ modes, respectively) [32,33]. The characteristics of the spectrum from the region altered by the THz pulses are identical to those of the crystal. Figure 2(b) shows the corresponding two-dimensional map of Fig. 1(g), plotting the intensity ratio between the Raman peaks at 105 and 148 cm$^{-1}$. The clear difference between the amorphous and crystalline phases means that the crystalline phase coincides with the region where the reflectivity change occurred. The growth speed in Figs. 1(d)–1(g) was estimated to be a few nanometers per pulse (5 μm for 5000 pulses), and it proceeded along the field direction.

To understand the processes occurring on an ultrafast time scale due to illumination with THz pulses and the resulting elongated crystal growth observed in Figs. 1(d)–1(g), the transient reflectivity change, $\Delta r/r$, within the crystallized area was measured using a pump–probe technique. The probe light pulse had a wavelength of 800 nm and a beam spot diameter of ≈ 1 μm [34]. The $\Delta r/r$ signal obtained from the crystalline phase at point

B in Fig. 1(g) is shown in Fig. 3(a) for three different incident electric field intensities. To prevent changes in the sample structure, the measurements were carried out using field strengths ($E_o \leq 150$ kV/cm) below those required for crystallization. Note that there was no difference between signals obtained in measurements using a ramp–up or ramp–down in THz intensity, as shown in the inset of Fig. 3(a); *i.e.*, the process was reversible.

For the largest field value, 150 kV/cm [in Fig. 3(a); red curve], the reflectivity dropped rapidly within the first few picoseconds and then recovered. This first component of the signal was linearly proportional to the square of the electric field (intensity) and thus shows the increase in absorption due to the electro-absorption effect [35]. A second drop in reflectivity with a delayed onset occurred at a delay of ≈ 25 ps, and the reflectively subsequently rose gently with a decay time of ≈ 500 ps (the heat diffusion length estimated from this trend was 4.5 nm [36]). The slow component corresponds to an increase in lattice temperature due to electron–phonon coupling, suggesting that the intensity at $t = 25$ ps is proportional to the total temperature rise $\Delta T_L$. This rise time of 25 ps was determined from the electron–phonon coupling coefficient $G$, which is $1.7 \times 10^9$ W cm$^{-3}$ K$^{-1}$ according to Fig. 3(a) [40].

The temperature rise $\Delta T_L$ can be expressed in terms of the electrical conductivity $\sigma$ (the Joule heat $\sigma E_i^2$ generated by the internal electric field $E_i$) and the specific heat of the lattice $C_L$, which is 1.25 J/(cm³·K) for the crystalline phase and 1.33 J/(cm³·K) for the amorphous phase [41]:

$$\Delta T_L = \frac{1}{C_L} \int \sigma(t) E_i(t)^2 dt . \tag{1}$$

The relationship between $\Delta T_L$ and Joule heating can be examined from that of the amplitude of the reflectivity change at 25 ps, $\Delta r_{\max}/r$ (which is proportional to $\Delta T_L$), and $E_o^2$ [Fig. 3(b)]. If the electrical conductivity is constant, Eq. (1) predicts that $\Delta T_L$ should increase linearly with $E_o^2$ (dashed line), which is inconsistent with the experimental data. This suggests that the electrical conductivity of the crystalline phase increases upon strong THz excitation. Upon illumination with a THz pulse, the initial crystallization starts around the antenna edge solely through the Joule heat generated by the large electric field in this region, which was confirmed experimentally (crystallization usually started near the corners of the antenna in our experiment) [31,42]. Nonetheless, the subsequent step-wise elongated crystal growth parallel to the field cannot be explained by the heat distribution that is obtained from Fig. 1(c).

The observed increase in $\sigma$ of the crystalline area probably originates from an increase in the electron density $N$. In particular, this suggests that Zener tunneling is responsible for the electron excitation in the crystalline phase; *i.e.*, when a strong electric field $E_i$ is applied to a semiconductor with a small bandgap, the band is modulated and electrons tunnel directly from the valence band into the conduction band. The number of electrons excited into the conduction band per unit volume and time can be expressed as [45-47]

$$N_{Zener}(E_i(t)) = \frac{e^2 E_i(t)^2 m_\mu^{1/2}}{18\pi\hbar^2 E_g^2} \exp\left(\frac{-\pi m_\mu^{1/2} E_g^{3/2}}{2\hbar e |E_i(t)|}\right), \tag{2}$$

where $\hbar$ is the Dirac constant, $e$ the elementary charge, $m_\mu$ the reduced mass (expressed as $m_\mu = m_e m_h/(m_e+m_h)$ using the effective masses $m_e$ and $m_h$ of electrons and holes), and $E_g$ the bandgap energy. We assumed that Eq. (2) holds for all time periods and determined the electron density in the crystal $N$ for a homogeneous distribution of additional electrons throughout the crystal volume at all time steps of the FDTD calculation (see the Supplemental Material [47]). Here, each cell in the FDTD model employs the local value of $E_i$ in order to account for the distribution of the electric field enhancement in the crystal phase that is induced by $E_o$. The electrical conductivity $\sigma$ can be determined by assuming that it is proportional to $N$ in accordance with the Drude model. In turn, $\Delta T_L$ can be calculated by inserting the obtained $\sigma$ and $E_i$ at the probe point into Eq. (1). Setting $m_e = 0.4m$ [48], $m_h = 0.2m$ [49] ($m$ is the electron rest mass), and $E_g = 0.24$ eV in Eqs. (1) and (2), we obtained a curve that reproduces the experimental results well (red solid line in Fig. 3(b)).

The relationship between $\Delta r_{max}/r$ and $N$ can be quantitatively examined using the THz-pump data with the signal obtained for 400-nm-pulse excitation (the yellow curve in Fig. 3(a); the calculation yields an optically excited electron density of $N_{opt} = 2 \times 10^{20}$ cm$^{-3}$ [56]). $\Delta r_{max}/r$ was $2.6 \times 10^{-2}$ at 25 ps for $E_o = 150$ kV/cm [Fig. 3(a); red curve], which is ten times larger than the $\Delta r_{max}/r$ for optical excitation ($\approx 2.8 \times 10^{-3}$), implying that the carrier density generated by the THz pulse is on the order of $10^{21}$ cm$^{-3}$. We obtained $N = 0.67 \times 10^{21}$ cm$^{-3}$ from a calculation based on Eq. (2) for $E_o = 150$ kV/cm, which is quantitatively consistent with the experimental result ($10^{21}$ cm$^{-3}$). Thus, for the electric field range where crystal growth occurs ($E_o = 175$ kV/cm), $N$ is larger than $10^{21}$ cm$^{-3}$ and thus $\sigma \geq 1 \times 10^3$ S/cm.

To clarify the distribution of the effective electric field during crystal growth, we used the above value of $\sigma = 1 \times 10^3$ S/cm for the crystalline region and performed a FDTD calculation. The results of the calculation [Fig. 3(c)] show that a highly concentrated field exists at the upper edge of the crystalline area, with values about 20 times that of the electric field acting on the amorphous region. Since the enhancement at the upper crystal edge is strong enough to exceed the crystallization temperature ($\Delta T_L = 130$ K [42]), we infer that crystal growth occurs along the electric field directions [see Figs. 1(e) and 1(f)] as a consequence of this concentration. Furthermore, if Zener tunneling is absent (*i.e.*, for the case of $\sigma = 100$ S/cm), the electric field enhancement factor at the edge of the crystalline area is only 8 times and $\Delta T_L = 26$ K, which suggests that the concentration of the electric field due to Zener tunneling induces the crystal growth in the region between the antennas.

To verify the dimensionality of the crystallization in Fig. 1, we compared the pulse number dependence of the crystal volume fraction $f(P)$ with the Johnson-Mehl-Avrami-Kolmogorov (JMAK) model [58,59]. We parameterized $f(P)$ with the reflectivity change, $\Delta R/R$, at points A, B, and C in Fig. 1(g) and plotted the corresponding normalized crystal volume fraction $f$ as a function of $P$ in Fig. 4. When crystallization starts after $P_i$ pulses at point $i$, the JMAK model leads to an Avrami equation of the form $f(P) = 1-\exp[-K(P-P_i)^n]$, with $K$ and $n$ being constant because the time for crystallization per pulse is constant ($P$ is the pulse number). Here, $n$ is the Avrami constant from which one can estimate the dimensionality of crystal growth. If $1 < n < 2$, the crystal grows one-dimensionally, while for $2 < n < 3$ it grows two-dimensionally [60]. Crystallization begins at $P_A$, $P_B$, and $P_C$, which define the time required for the crystalline nuclei to reach the critical radius [61], and $\Delta R/R$ saturates at $P_f$. $f(P)$ was fitted by varying $n$; the fitting confirmed that $n_i$ changes dramatically when a threshold, defined by $P_m$, is reached (for $P = P_m$, the crystalized area in the image connects the upper and lower Au antennas). The region before $P_m$ can be classified as one-dimensional growth, since growth proceeds with $n < 2$ as illustrated in the inset of Fig. 4. On the other hand, $n$ (= 2.48) in the region after $P_m$ is consistent with crystallization driven by pulsed laser illumination ($n = 2.5$ [60]), indicating that it proceeds via isotropic heating of the entire crystal volume.

In conclusion, the observed nonlinear increase in the conductivity of the crystallized region is a result of Zener tunneling; the tunneling leads to local heating which causes a one-dimensional crystal growth. Besides revealing the carrier dynamics in GST under the influence of electric fields that are relevant to crystal growth, our results indicate that the observed growth rate was a few nm per pulse, which provides a sophisticated method for miniaturizing memory devices down to the nanometer scale. Finally, we note that the role of the nucleation in the crystallization process is still a matter of debate. As such, the characterization of the atomic structure is essential to clarify the crystal growth mechanism under high electric fields.

## Acknowledgements

xThis study was supported by PRESTO (No. JPMJPR1427) and CREST (No. JPMJCR14F1) grants from JST and the KAKENHI grant 17H06228 from JSPS. We are grateful to Naoki Yokoyama and Hiroyuki Akinaga for valuable comments.

**Figure captions**

**FIG. 1.** (a) Experimental setup. The Au antenna enhances the THz electric field over the 5 μm gap. (b) Measured temporal profile of the pulse. (c) Electric field enhancement and field profiles in the gap. White arrows indicate the electric field directions. (d)–(g) Microscope images of spatial reflectivity changes ($\Delta R/R$) after THz irradiation as a function of pulse number. The images were obtained by averaging over 450 pulses (225 pulses before and after the corresponding center pulse) and dividing the result with the data obtained without THz excitation.

**FIG. 2.** (a) Comparison of Raman spectra for point B in Fig. 1 (g) (red), crystalline GST (black) and amorphous GST (blue). Arrows indicate the mode positions, and the curves are offset for clarity. (b) Spatial map of intensity ratio of Raman signals at (i) 105 and (ii) 148 cm$^{-1}$. Since Au has a constant signal over a broad region, the ratio is about unity.

**FIG. 3.** (a) Time-resolved reflectivity change for point B [Fig.1 (g)] for different THz electric field strengths (black: $E_o = 50$ kV/cm; blue: $E_o = 100$ kV/cm; red: $E_o = 150$ kV/cm) and optical fluence of 0.12 mJ/cm$^2$ (yellow). (b) THz intensity dependence of peak reflectivity modulation $\Delta r_{max}/r$ at time delay of 25 ps in (a). No change was observed for the amorphous region because the electric field enhancement at the probe spot is significantly lower than the electric field enhancement near the antenna. While the latter reaches the crystallization temperature, the temperature of amorphous region at the probe spot increases only by about 3 K [42,62]. The scale of $\Delta T_L$ is on the right. The data was measured by varying $E_o$ as follows: 150, 120, 100, 80, 50, 90, 110, 130, 60, 70, and 140 kV/cm. Because the data line up on a single curve, the sample condition was not altered during the measurements and repeatability is confirmed. (c) Electric field enhancement in the 5-μm gap during the actual crystal growth indicated in Fig. 1(e). The electric field

is dramatically enhanced at the edge of the crystalline area with a conductivity of 1000 S/cm.

**FIG. 4.** THz pulse number dependence of crystal volume fraction $f(P)$ obtained experimentally at different locations marked as points A (red square), B (green square), and C (blue square) in Fig. 1(g). The experimental $f(P)$ is obtained by dividing the reflectivity $\Delta R(P)$ by the saturated maximum value: $\Delta R_{max} = \Delta R(P = 20000)$. The red, green, and blue solid curves are fittings of the Avrami equation to the experimental values at points A, B, and C ranging from $P_A$, $P_B$, and $P_C$ to $P_m$. The black curve is a fitting using the averaged data measured at the points A, B, C for $P > P_m$. $P_A = 1.1 \times 10^4$, $P_B = 1.2 \times 10^4$, $P_C = 1.4 \times 10^4$, $P_{avg} = 1.5 \times 10^4$, $K_A = 2.0 \times 10^{-4}$, $K_B = 6.6 \times 10^{-5}$, $K_C = 1.5 \times 10^{-6}$, $K_{avg} = 1.1 \times 10^{-8}$, $n_A = 0.90$, $n_B = 1.05$, $n_C = 1.67$, $n_{avg} = 2.45$. Inset: Mechanism of the crystal growth under applied electric field.

**FIG. 1.**

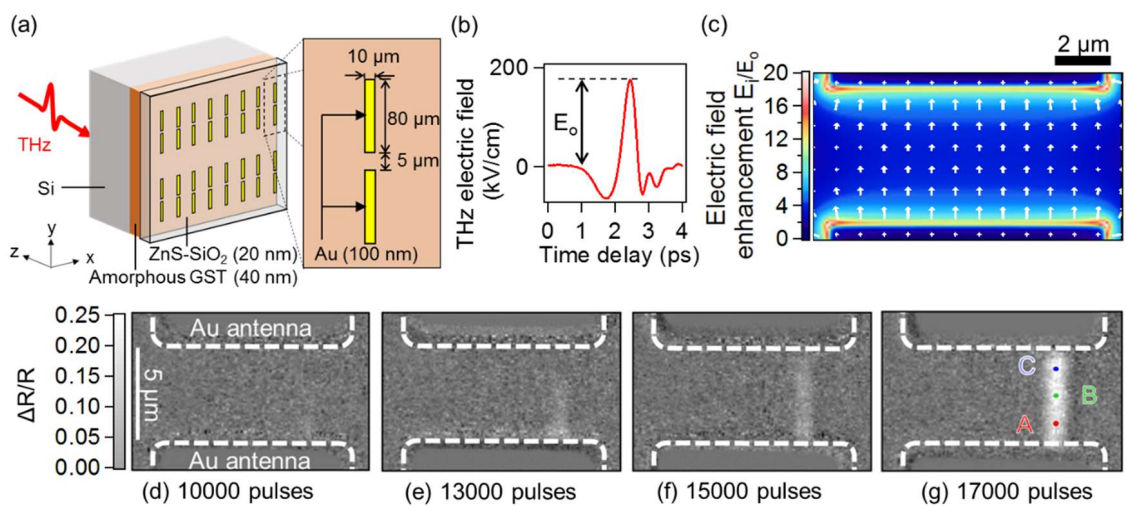

Y. Sanari

**FIG. 2.**

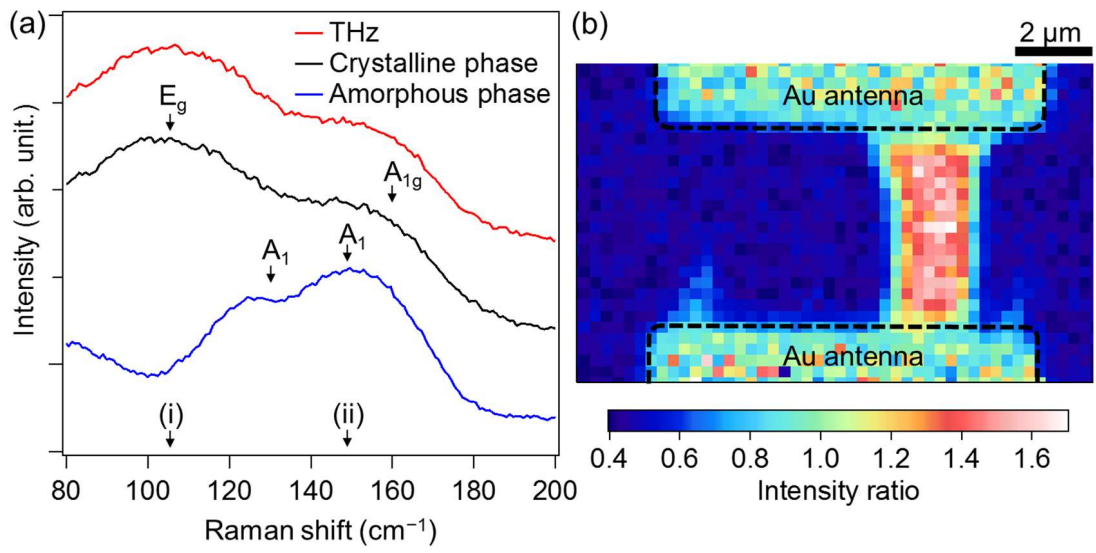

Y. Sanari

**FIG. 3.**

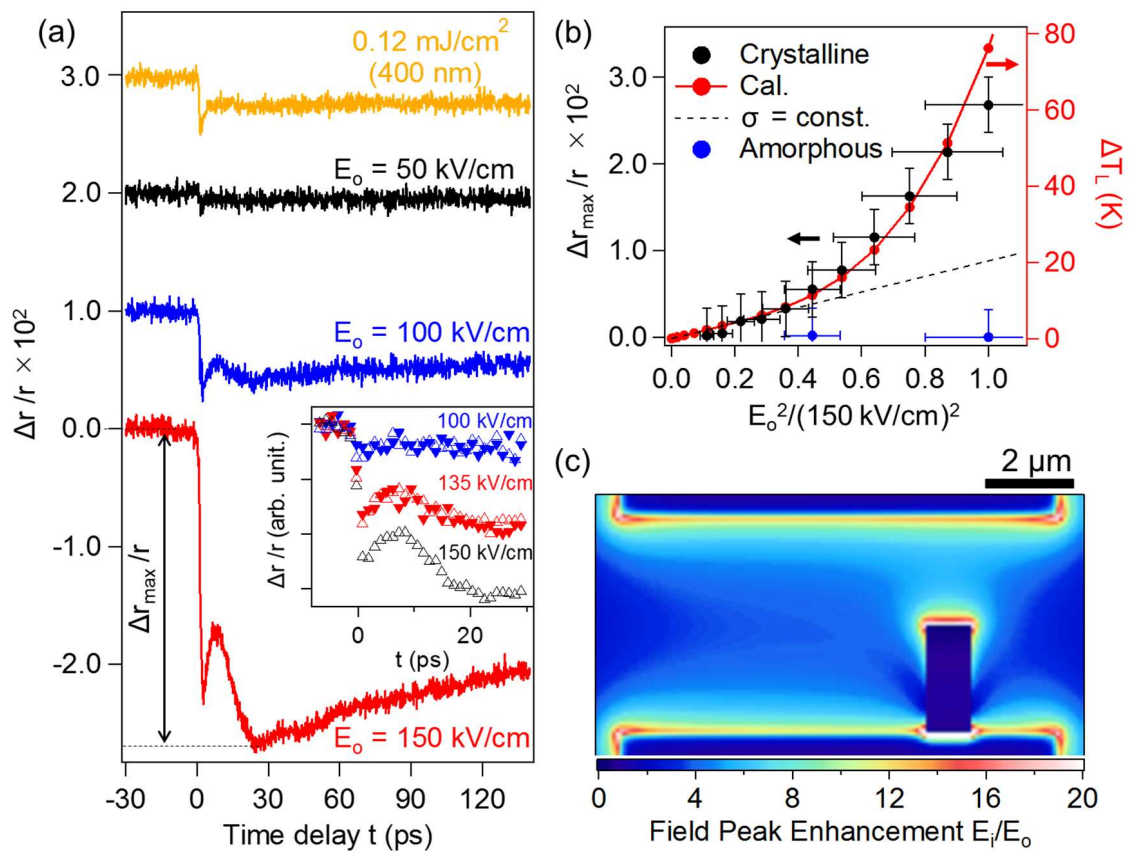

Y. Sanari

**FIG. 4.**

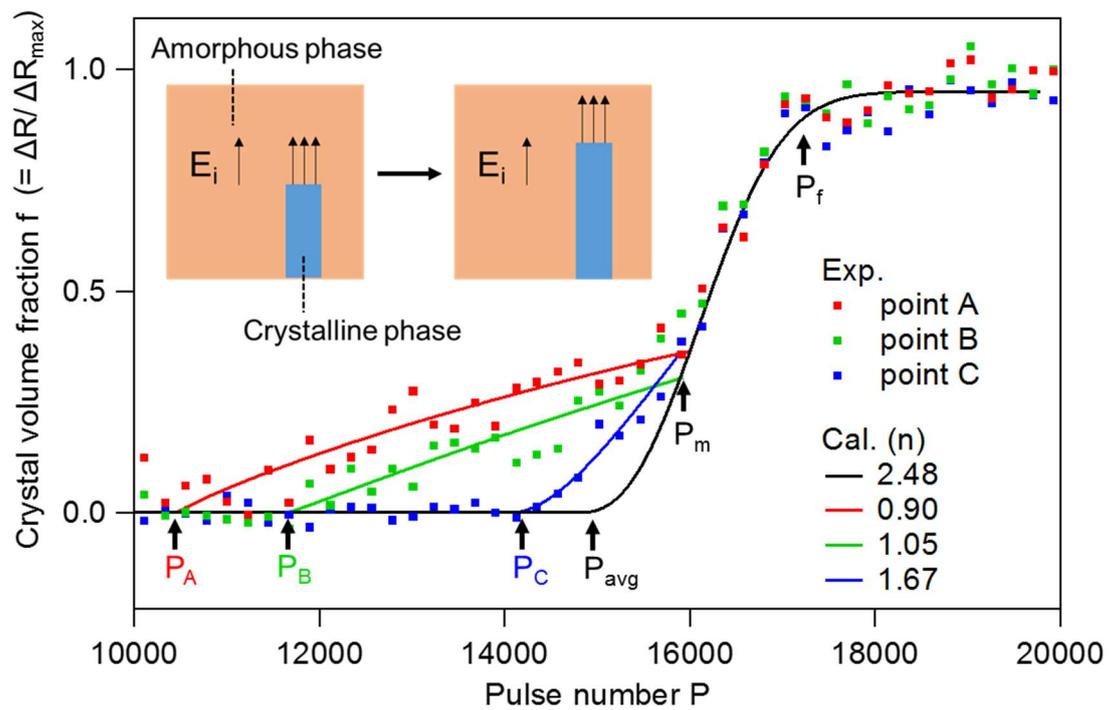

Y. Sanari